\documentclass{article}
\usepackage{graphicx}
\usepackage{amsmath}
\usepackage{amsfonts}
\usepackage{amssymb}
\begin{document}

\title{ Games with Quantum Analogues}
\author{J. M. A. Figueiredo\\Universidade Federal de Minas Gerais - Dept. de F\'{i}sica \\Caixa Postal 702 - Belo Horizonte - 30.123-970\\Brazil\\email: josef@fisica.ufmg.br}
\maketitle
\begin{abstract}
A new class of stochastic variables, governed by a specifice set of rules, is
introduced. These rules force them to loose some properties usually assumed
for this kind of variables. We demonstrate that stochastic processes driven by
these random sources must be described using an probability amplitude
formalism in a close resemblance to Quantum Theory. This fact shows, for the
first time, that probability amplitudes are a general concept and is not
exclusive to the formalism of Quantum Theory. Application of these rules to a
noisy, one-dimensional motion, leds to a probability structure homomorphic to
Quantum Mechanics.

Stochastic Models, Game Theory, Quantum Theory
\end{abstract}

The understanding that noise drives a wide variety of phenomena in Nature
brought to many branches of Science the conviction that randomness is a
fundamental property that must be considered in any real modeling of the
physical world. In fact, all theories yet developed that include noise as a
element driving an otherwise deterministic process still remain making use of
this concept even when specific changes on its mathematical modeling may
eventually be refined. The only exception is Quantum Theory, which makes use
of probabilistic tools without specific reference to any noise source. In its
early developments Einstein, Hopf and Stern \cite{miloni} tried to explain
some electrodynamic phenomena using a noisy electromagnetic field, in addition
to the usual determinist one, in the dynamical equations describing particle
motion. However, later developments of the Quantum Theory led to a noiseless
model, that acts as a noise source but not depending, on its fundamental
epistemological components, of a noise source. Despite this very important
exception we can assert, without doubt, that stochastic physical theories are
epistemologically robust and confident.

It is clear, from the above arguments, that two general classes of probability
theories exist. The classical ones makes use of probabilities or probability
densities and is used to describe stochastic processes. On the other hand
quantum probabilities must be determined from a probability amplitude and do
not describe a stochastic process. In fact quantum probability is considered a
intrinsic property of Nature; this way it is strongly believed that
probability amplitudes are exclusive to quantum phenomena which drives any
other macroscopic manifestation of physical phenomena. No other manifestation
of statistical phenomena, in any field, may have, in principle, the
characteristics of the quantum probability. That is the point we wanto to
discuss in this work. We present a new model for the random variable itself
thus affecting all existing stochastic theories whenever the novelties
introduced here are applicable. Stochastic processes driven by this kind of
variable make use of a probability amplitude formalism even thought no
reference to quantum phenomena is made.

Usual stochastic processes present two main components: a determinist law plus
an additional stochastic term depending on the value of a random variable.
This variable assumes its values in a unpredictable way but once its
instantaneous fluctuating value is defined the system responds to it in a
predicable way. No doubt concerning the ''reading'' process is introduced. A
causal relation exists between the actual value assumed by the random variable
and the one in fact used by the system in its deterministic response to noise.
This reading process then has a property, borrowed from similar concept, usual
in Quantum Theory, we call ''realism'' because the actual value of the random
variable may, at least in principle, be known.

We want to deny the necessity of realism in a stochastic theory by introducing
the concept of \emph{incomplete} random variable. A system responds to this
kind of variable in a two steps process. In the first one the variable
fluctuates in the usual way but its value is not readily available to the
system that must read it. In this second step some errors may occur. Two types
of errors will be considered here. They are as defined by the Q-rules set
postulated below:

\begin{itemize}
\item  rule one: the fluctuating value is not read; the fluctuation is lost.
System's state is frozen until the next round. A clear breakdown of causality
is introduced at this point.

\item  rule two: some readings are wrong. Another value for the fluctuating
variable, not the real one, is used by the system to drive its dynamics. The
value is lost but not the reading.
\end{itemize}

Before a explicitly mathematical formalization of these rules it is worth to
stress that all them are algorithmic even though they break causality and
realism. In fact, we can't assert to an incomplete random variable any
criteria of realism because it is impossible to know its exact value; at the
same time the relation between the system and the noise source is not causal
because nobody can guarantee system's dynamics will be dictated by a given
fluctuation. In some sense, however, reality cannot be completely discarded
because fluctuations itself are real. Notwithstanding, realism, as defined
above, definitively does not exist. More formally, let us consider a discrete
random variable $Y$ that can have $M$ values, and its associated probability
distribution $P\left(  y\right)  $. For definiteness assume that $P\left(
y_{j}\right)  >0$ for all $j\in\left\{  1,2,...M\right\}  $. If $Y$ is a
incomplete random variable then its value is not directly accessible to the
system. The actual histogram for $Y$ is not any more given by $P\left(
y\right)  $. Instead, a modified one that takes into account the Q-rules must
be used. This modified probability distribution, the non-classical one, will
be written as $\textsl{p}_{j}=P_{j}+C_{j}$ where $P_{j}\equiv P\left(
y_{j}\right)  $ and $C_{j}$ is the non-classical term. This term has
contributions from both Q-rules. From the first one we have the probability of
lost readings given by $\gamma_{j}P_{j}$. Thus, $P_{j}\left(  1-\gamma
_{j}\right)  $ is the probability of successful readings. Two kinds of
contributions arise from rule two. The probability of reading state $j$ may be
enhanced by a wrong reading of another state; on the other hand this
probability may be decreased by a wrong reading that converts state $j$ to
another one. If $\Gamma_{j,l}$ $\left(  j\neq l\right)  $ is the probability
that a realization giving the value $y_{l}$ be read as $y_{j}$ then we write
rule two as $\sum_{l=1,l\neq j}^{M}\left(  \Gamma_{j,l}P_{l}-\Gamma_{l,j}%
P_{j}\right)  $. Defining $\tilde{C}_{j,l}\equiv\Gamma_{j,l}P_{l}-\left(
\Gamma_{l,j}+\frac{\gamma_{j}}{M-1}\right)  P_{j}$ $\left(  j\neq l\right)  $
we get the histogram effectively read for variable $Y $%
\begin{equation}
\textsl{p}_{j}=P_{j}+\sum_{l=1,l\neq j}^{M}\tilde{C}_{j,l} \label{Q-histog}%
\end{equation}
Observe that the non-classical terms $\tilde{C}_{j,l}$ may have any sign;
consequently they cannot be seen as a truly probability term. Only the non
classical distribution $\textsl{p}$, that possesses the desired positiveness,
is accessible to the system. That means we cannot write $p$ as the sum of two
independent process, one driven by $P$ and the other by the non-classical
terms. At this point we explicitly break realism because the observed
histogram corresponds to $p$, not to $P$. So we cannot directly measure or
observe the noise source. This way we see that incomplete random variables
present a simple algorithm in their realization, although their conception are
more complicated than the usual ones. For instance, consider a roulette with
$M$ stops and an imperfect camera that reads pointer's position at each round.
The roulette drives a game but pointer's position is only accessible to
players through the camera that may generate wrong readings (or no readings at
all), according to the Q-rules. The roulette is real but this reality is only
conceptual, since the actual value used in the game comes from camera readings.

We are now in position to study a stochastic process driven by a incomplete
random variable. Let us initially consider a discrete time process (a game)
\cite{note}, of the following general form
\begin{equation}
f\left(  x_{n+1},x_{n}\right)  =g\left(  x_{n}\right)  w_{n} \label{game}%
\end{equation}
where $W$ is the (uncorrelated) stochastic variable and $\varphi\left(
w\right)  $ its associated probability distribution. For $g\left(  x\right)
=0$ the game evolves according a deterministic rule, defining player's
strategy. A noise source includes a chance ingredient, that makes best
strategy choices non-trivial. We regard eqn$\left(  \ref{game}\right)  $ as
defining the value the stochastic variable $x_{n+1}$ will take at round $n+1$,
considering that it takes the value $x_{n}$ at round $n$. The joint
probability density $\Pi\left(  x_{N},x_{N-1},....,x_{1}\right)  $ completely
describe this process and for a given sampling sequence of $W$ it is given by
\cite{kampen}:
\[
\Pi\left(  x_{N},x_{N-1},....,x_{1}\right)  =\int\prod_{n=1}^{N}\delta\left(
f\left(  x_{n+1},x_{n}\right)  -g\left(  x_{n}\right)  w_{n}\right)
\varphi\left(  w_{n}\right)  dw_{n}
\]
which in the discretized model assumes the form
\begin{equation}
\Pi\left(  x_{N},x_{N-1},....,x_{1}\right)  =\sum_{j1=1}...\sum_{jN=1}%
\prod_{n=1}^{N}\delta\left(  f\left(  x_{n+1},x_{n}\right)  -g\left(
x_{n}\right)  w_{jn}\right)  \varphi\left(  w_{jn}\right)  \label{paths}%
\end{equation}
The product $\prod_{n=1}^{N}\varphi\left(  w_{jn}\right)  $ defines a ordered
sampling sequence $\left\{  w_{j1}...w_{jN}\right\}  $ of the bare random
variable $W$ which we will call a ''path''; it is easy to see that in the
above summation there are $M^{N}$ paths. Path concept in stochastic processes
was used initially by Onsager and Machlup \cite{onsager} in order to describe
the effect of fluctuations on non-equilibrium systems. Later Graham
\cite{graham} developed a more general formalism to include sophisticated
dynamics satisfying the Liouville equation which successfully allowed works in
systems subject to colored noise \cite{colored}. We shall adopt eqn$\left(
\ref{paths}\right)  $ as a strong support to later developments in the present work.

Now it is natural to introduce the effects of the Q-rules on stochastic
processes by imposing that $W$ be an incomplete random variable $Y$ and
studying how incompleteness affects eqn$\left(  \ref{paths}\right)  $.
Histogram of $Y$ is given by eqn$\left(  \ref{Q-histog}\right)  $ and at round
$n$ it assumes the value $y_{jn}$. Then to any path there is an associated
product of the form
\begin{equation}
\prod_{n=1}^{N}\textsl{p}\left(  y_{jn}\right)  =\prod_{n=1}^{N}\left(
P_{jn}+\sum_{l=1,l\neq j}^{M}\tilde{C}_{jn,l}\right)  \label{qpaths}%
\end{equation}
which, for a incomplete variable, branches in a sum having $M^{N}$ terms, as
seen from the left side of the above equation. There are $M^{N}$ paths so the
sum over all paths, involved in eqn$\left(  \ref{paths}\right)  $, has
$M^{2N}$ terms. The set of all paths, $\Xi$, may be represented as the union
of two disjoint subsets: $\Xi=\Omega\bigcup\Lambda$ where $\Omega$ is the set
of all paths containing only $P$-terms and $\Lambda$ contains elements with at
least one non-classical contribution. Thus $\Omega$ consists in the set of all
''classical'' paths, those valid for a complete random variable, and has
$M^{N}$ elements.

In sequence we make a simplification in the model, assuming that
\begin{equation}
\tilde{C}_{j,l}=\tilde{C}_{l,j}\Longrightarrow\tilde{C}_{j,l}=-\frac
{\gamma_{j}P_{j}+\gamma_{l}P_{l}}{2\left(  M-1\right)  }\leq0 \label{symm}%
\end{equation}
This means that vacuum losses $\left(  \gamma\right)  $ completely dominate
the chance of non-classical effects during the reading process. The number of
elements in $\Lambda$ is reduced due to this symmetry as shown below. Writing
$M^{2N}=\left(  M^{2}-M+M\right)  ^{N}$ we get
\[
M^{2N}=\sum_{l=0}^{N}\binom{N}{l}M^{\left(  N-l\right)  }\left(
M^{2}-M\right)  ^{l}
\]
showing that there are $\binom{N}{l}M^{\left(  N-l\right)  }\left(
M^{2}-M\right)  ^{l}$paths with $l$ non-classical contributions. The symmetry
expressed by eqn$\left(  \ref{symm}\right)  $ demands this number be reduced
to $\binom{N}{l}M^{\left(  N-l\right)  }\left(  \frac{M^{2}-M}{2}\right)
^{l}$ terms, generating a total number of independent non-classical paths
equal to
\[
\sum_{l=1}^{N}\binom{N}{l}M^{\left(  N-l\right)  }\left(  \frac{M^{2}-M}%
{2}\right)  ^{l}=\left(  \frac{M^{2}+M}{2}\right)  ^{N}-M^{N}
\]
caused by the existence of twin paths, those having labels in the
non-classical contributions exchanged. This result reduces the number of
elements in $\Xi$ to $\left(  \frac{M^{2}+M}{2}\right)  ^{N}$.

We will now show that classical paths may index all non-classical ones as
well, in a very specific way. To this end we rewrite
\begin{equation}
\tilde{C}_{j,l}=-\frac{\sqrt{P_{l}P_{j}}}{2\left(  M-1\right)  }\left(
\gamma_{l}\sqrt{\frac{P_{l}}{P_{j}}}+\gamma_{j}\sqrt{\frac{P_{j}}{P_{l}}%
}\right)  \equiv\sqrt{P_{j}}\sqrt{P_{l}}d_{j,l} \label{symm1}%
\end{equation}
and $P_{j}=\sqrt{P_{j}}\sqrt{P_{j}}$, in such a way that each non-classical
path has the structure of a product of $2N$ factors of the type $\sqrt{P}$,
followed by a product of $d^{\prime}s$, in the same number of the existing
non-classical terms in the considered path. Note that classical paths can also
be rewritten as a sequence of type $\sqrt{P}$ having, obviously, no $d$-terms.
Since all $d%
%TCIMACRO{\UNICODE{0xb4}}%
%BeginExpansion
\acute{}%
%EndExpansion
s$ are negative the value of a non-classical path may have any sign. In
considering this type of sequences, another marked difference between elements
in $\Omega$ and $\Lambda$ is that each non-classical contribution has
necessarily cross terms like $\sqrt{P_{j}}\sqrt{P_{l}}$, with $l\neq j$.
However, note that non-classical paths may have classical-like segments with
the same structure of elements in $\Omega$. We call \emph{radix} $\left(
R\right)  $ of a sequence its non-$d$ terms part. That is, the radix is a
sequence of terms like $\sqrt{P_{j}}\sqrt{P_{l}}$ (for all $l$ and $j$).
Consequently all elements in $\Lambda$ have the structure $R\prod d$. The
product runs over the number $L_{path}$ of non-classical terms in the
considered sequence. Thus we define an auxiliary set $\Delta$ composed by all
possible radices. It's easy to see that $\Omega$ is a subset of $\Delta$ which
have the same number of elements of $\Xi$. In what follows we shall use these
facts in order to construct a unified structure for that set.

Let us consider the set $\tilde{H}$ of all classical-like paths of type
$\sqrt{P}$; that is, paths in $\tilde{H}$ are just those sequences in $\Omega$
where $P^{\prime}s$ are substituted by the corresponding $\sqrt{P^{\prime}s}$.
Thus each element $R_{\tilde{H}}$ in $\tilde{H}$ may be written as
$R_{\tilde{H}}=\sqrt{R_{\Omega}}$, where $R_{\Omega}$ is the corresponding
path in $\Omega$. On the other hand since $\Delta$ consists of all
combinations of ordered sequences of $\sqrt{P}$-type we may write each element
there as a cross-product of some elements in $\tilde{H}$. That is, we always
can write each element $R_{\Delta}$ in $\Delta$ as $R_{\Delta}=R_{\tilde{H}%
}R_{\tilde{H}}^{\prime}$ for two carefully chosen elements in $\tilde{H}$. As
a result we write the important result that $\Delta=\tilde{H}\otimes\tilde{H}%
$. Now we define an extended set $H$ by assigning to each element in
$\tilde{H}$ a ``phase'', a complex number $\exp(i\varphi_{\Omega})$, in such a
way that the $M^{N}$ elements in $H$ have the general form $S_{H}=R_{\tilde
{H}}\exp(i\varphi_{\Omega})=\sqrt{R_{\Omega}}\exp(i\varphi_{\Omega})$.
Consequently, since $S_{H}\left(  S_{H}\right)  ^{\ast}=R_{\Omega}$, we see
that $\Omega\supset H\otimes H^{\ast}$. On the other hand the radix of an
element in $H$ is just the corresponding element in $\tilde{H}$. This way we
see that elements of $H\otimes H^{\ast}$ have the general form $R_{\Delta}%
\exp(i\left(  \varphi_{\Omega}-\varphi_{\Omega^{%
%TCIMACRO{\UNICODE{0xb4}}%
%BeginExpansion
\acute{}%
%EndExpansion
}}\right)  )$ or $R_{\Delta}\exp(-i\left(  \varphi_{\Omega}-\varphi_{\Omega^{%
%TCIMACRO{\UNICODE{0xb4}}%
%BeginExpansion
\acute{}%
%EndExpansion
}}\right)  )$. The sum $\tilde{\Xi}=$ $\sum H\otimes H^{\ast}$ has
$M^{N}+M^{N}\left(  M^{N}-1\right)  /2$ terms (all them real numbers); out of
these there are $M^{N}$ terms having no phase contribution. They correspond to
the classical paths belonging to $\Omega$; the remaining ones can be collected
into $\left(  \frac{M^{2}+M}{2}\right)  ^{N}-M^{N}$ independent radices
$R_{\Delta}$ times a linar combination of phase terms, each one expressed as a
$2\cos\left(  \varphi_{\Omega}-\varphi_{\Omega^{%
%TCIMACRO{\UNICODE{0xb4}}%
%BeginExpansion
\acute{}%
%EndExpansion
}}\right)  $. This last term is symmetric under exchange of argument indices,
in as much as eqn$\left(  \ref{symm1}\right)  $is. Hence the same argument
used to count the twin paths is applicable. This reduces the number of terms
in $\tilde{\Xi}$ to $\left(  \frac{M^{2}+M}{2}\right)  ^{N}$ which is the same
number as the sum over all paths in $\Xi$. The above results allow us to write
the main result of this work:
\begin{equation}
\sum_{\Xi}=\left|  \sum_{H}\right|  ^{2} \label{summ}%
\end{equation}
valid if we make the association
\begin{equation}
\sum_{L_{path}}\prod^{L_{ph}}d_{ji}\equiv\sum_{L_{path}}\prod^{L_{ph}}%
\cos\left(  \varphi_{i}-\varphi_{j}\right)  \label{coss}%
\end{equation}
where $i$ and $j$ are classical paths used to index the corresponding common
radix $R_{\Delta}$ of elements in $\tilde{H}$. That is the sum over all paths
in $\Xi$ may be written as a squared sum over all paths in $H$. In analogy to
quantum theory we call amplitudes the elements in $H$. Thus, in order to sum
up the right result in $\Xi$ sample paths in $H$ (which consists of classical
paths for the amplitudes) must ''interfere''; that is, in this space, paths,
not single realizations of the random variable, are the basic objects. Notice
that the phase of a path is a non-local object in the sense we cannot assign
to it a single specific process, once it is defined for combinations of
$d$-terms. In fact, even when $Y$ is uncorrelated, the phase depends on the
whole sequence in a path. In this space realism breaks down since amplitudes
cannot be observed in any particular realization of $Y$. In the same way it
will be shown below that causality is not present as well. Therefore, in the
context of this work probability amplitudes describe in a unified way the
whole effects of incomplete random variables in such way that interference of
these amplitudes means that cross-effects on probabilities due to the Q-rules
are relevant.

In order to obtain values for the $M^{N}$ phases we expand them in a
(truncated) path-dependent power series, each one having a number of terms
equal to the integer part of $\left[  \left(  M+1\right)  /2\right]  ^{N}$, as
follows
\begin{equation}
\varphi_{l}=\varphi_{N}+\sum_{nl=1}^{N}A_{nl}y_{nl}+\sum_{nl=1}^{N}\sum
_{ml=1}^{N}\tilde{B}_{nl}^{ml}y_{nl}y_{ml}+... \label{series}%
\end{equation}
where $l\in\left[  1,M^{N}\right]  $ and $\varphi_{N}$ is zero if $M^{N}$ is
odd. The sequence of values the random variable takes in a given path may be
rescaled by incorporation of the linear coefficients $\left\{  A_{nl}\right\}
$ once a similar renormalization in the high order coefficients is
consistently performed. This way we get a new set of (path-dependent) values
for $Y$. This new set is obtained by linearization of eqn$\left(
\ref{coss}\right)  $ and does not represent any restriction on the formulation
of the problem since, as seen from eqn$\left(  \ref{paths}\right)  $, their
elements are dummy variables. As a result it is possible to rewrite
eqn$\left(  \ref{series}\right)  $ as
\begin{equation}
\varphi_{l}=\varphi_{N}+\sum_{nl=1}^{N}\tilde{y}_{nl}+\sum_{nl=1}^{N}%
\sum_{ml=1}^{N}B_{nl}^{ml}\tilde{y}_{nl}\tilde{y}_{ml}+.. \label{series1}%
\end{equation}
where now it is implicit that the set of sampling variables $\tilde{y}$ is
path-dependent. For each path we have chosen the truncation schema carefully
by collecting exactly $\left[  \left(  M+1\right)  /2\right]  ^{N}$
coefficients in the above series. This sums up to a set of $M^{N}\left[
\left(  M+1\right)  /2\right]  ^{N}$ elements in the whole set $H$. Then just
make use of the same number of equations displayed in eqn$\left(
\ref{coss}\right)  $ to solve for these coefficients, completing this way the
construction of $H$. Now we have at hand a space consisting of classical
paths, homomorphic to $\Omega$, but whose elements are probability amplitudes.
These amplitudes are the mathematical objects used to treat incomplete random variables.

To the phase of a path it is not possible to assign any single realization of
the stochastic variable. However it is possible to assign a phase to a segment
of a path although single-event association still remains invalid. To define
the phase of a segment we rewrite eqn$\left(  \ref{series1}\right)  $ as a sum
of $N$ terms:
\begin{equation}
\varphi_{l}=\sum_{n=1}^{N}\left[  \frac{\varphi_{Nl}}{N}+y_{nl}\left(
1+\sum_{m=1}^{N}B_{nl}^{ml}y_{ml}+...\right)  \right]  \equiv\sum_{n=1}%
^{N}\phi_{nl} \label{phase}%
\end{equation}
allowing us to write an element in $H$ as $S_{H}=\prod_{n=1}^{N}\sqrt{P_{jn}%
}e^{i\phi_{nl}}$. That is, incomplete random variables must be described by a
probability amplitude
\begin{equation}
U\left(  j,n,N\right)  \equiv\sqrt{P_{j}}e^{i\varphi\left(  j,n,N\right)  }
\label{amplit}%
\end{equation}
in such a way that $\left|  U\left(  j,n,N\right)  \right|  ^{2}=P_{j}$. The
phase of this probability amplitude is process-dependent and cannot be
observed in any single realization of the variable, which always results on
its effective histogram $p$. In this case we say that the bare distribution
$P$ is \emph{hidden}. Furthermore, the phase value, for a given path, depends
on all realizations of $y$ that closes the considered path, including those
chosen at future rounds (in the considered path). This property breaks down
causality in the phase definition for individual realizations in the sense
that future rounds define the present. However this fact does not configure a
violation of causality for the whole process once the paths, which form the
basic blocks in constructing the transition amplitudes, are causal.

Returning back to the continuum, we must then describe an incomplete random
variable $y$ by a probability amplitude $U\left(  y,n,N\right)  =\sqrt
{P\left(  y\right)  }e^{i\varphi\left(  y,n,N\right)  }$ that must be
considered in any stochastic process it drives. We cannot measure the phase of
a probability amplitude in any single event measurement; thus it is not
possible to assign to it an element of reality. This is the main difference
from usual (classic) stochastic variables that are completely described by
their bare probability distribution $P\left(  y\right)  $. Now we are in
position of analyze how the Q-rules affect a specific game.

To do this let us consider the process defined in eqn$\left(  \ref{game}%
\right)  $. For any given initial state $E\left(  x_{0}\right)  $, the
probability distribution after $N$ rounds is \cite{gardner}
\[
E\left(  x_{N}\mid x_{0}\right)  =\int E\left(  x_{0}\right)  \prod_{n=0}%
^{N}\delta\left(  f\left(  x_{n+1},x_{n}\right)  -y_{n}\right)  P\left(
y_{n}\right)  dy_{n}dx_{n}
\]
but as written this equation describes the classical version of $Y$. If the
Q-rules apply up, the phase of the probability distribution should be
considered. So, as it stands this equation cannot be used unless we do make
the substitution $P\rightarrow p$; we also perform an integration over the
random variable $y$ that results in
\[
E\left(  x_{N}\right)  =\int E\left(  x_{0}\right)  \prod_{n=0}^{N-1}p\left(
f\left(  x_{n+1},x_{n}\right)  \right)  dx_{n}
\]
which is recognized as the Chapman-Kolmogorov equation for a Markov process
having transition probabilities given by $p\left(  f\left(  x_{n+1}%
,x_{n}\right)  \right)  $. After discretization this equation seems to have
the same path topology of eqns$\left(  \ref{paths}\right)  $ and $\left(
\ref{qpaths}\right)  $; however, an additional sum over the state at $n=0$ is
present. It can naturally be inserted in the paths if we define $E\left(
x_{0}\right)  \equiv\left|  \psi\left(  x_{0}\right)  \right|  ^{2}$ and use
eqn$\left(  \ref{summ}\right)  $ to get an analogous expression for the
probability distribution at time $N$ given by $E\left(  x_{N}\right)  =\left|
\psi\left(  x_{N}\right)  \right|  ^{2}$, where
\begin{equation}
\psi\left(  x_{N}\right)  =\sum_{H}\sqrt{P\left(  f\left(  x_{n+1}%
,x_{n}\right)  \right)  }e^{i\varphi\left(  path\right)  }\psi\left(
x_{0}\right)  \label{feymman}%
\end{equation}
which, after recovering the continuum, may be written as
\begin{equation}
\psi\left(  x_{N}\right)  =\int\psi\left(  x_{0}\right)  \prod_{n=0}%
^{N-1}\sqrt{P\left(  f\left(  x_{n+1},x_{n}\right)  \right)  }e^{i\varphi
\left(  path\right)  }dx_{n} \label{feymman1}%
\end{equation}
Notice that now the bare probability $P$ is used in place of $p$ as prescribed
by eqn$\left(  \ref{amplit}\right)  $. We arrived at a stochastic version of
the Feynman-Kac formula, generalized to any kind of (one-player) game. Note
that it is possible to control intensities of each kind of bare process
altering both the radix $\left(  \sqrt{P}\right)  $ and the phase $\left(
d^{\prime}s\right)  $ of a given path in the above cited equation.

We can use this equation in any Markovian process satisfying eqn$\left(
\ref{game}\right)  $ driven by incomplete random variables. Since the set of
rules defining this kind of variables are algorithmic, a new generation of
games may be defined and constructed, impinging new challenges to Game Theory.
These new games are not the same as the incomplete games \cite{game}, those
where players don't know about the decisions the others have taken. The use of
incomplete variables in a game makes the reading process imperfect but this
limitation does not deny the knowledge of the readings be shared by all
players. However, it may occur that players have their own reading apparatus.
In this case we have a truly incomplete non-classical game. In this case
decision theory and best strategy modeling must take into account the
incompleteness of the random source. The possibilities opened by this approach
should improve our comprehension about algorithmic probability amplitude
effects through modeling and implementation of Q-rules in any specific game.
Because we make use of probability amplitude formalism in these games it is
reasonable to call them Quantum Games. In this work we do not develop further
any general discussion about the consequences of the concepts presented here
on Game Theory. Our main interest here was solely to show the existence of
this class of games and how the probability amplitude formalism may arise in a
truly stochastic theory. However, it would be interesting to analyze some real
physical game since a similarity of eqn$\left(  \ref{feymman1}\right)  $ to
path integrals in Quantum Theory is unavoidable. To this end we present below
an important sample-game describing the motion of a classical particle subject
to a noisy environment, which is very similar to those processes considered in
ref \cite{colored}. Then we will show that a probability amplitude for this
game can be constructed in a way that resembles the properties of a quantum
particle. A remarkable point is the evidence shown here that probability
amplitude effects are not exclusive of Quantum Theory neither a mysterious
fundamental working mechanism in the Universe. In fact this formalism is now
trivial once infinity types of stochastic processes may use it where some
basic, algorithmic rules are the fundamental assumptions, not the use of the
probability amplitude formalism itself. This fact changes naturally our
attention in direction to Quantum Mechanics in order to look for a more
fundamental phenomena that justify its basics axioms. We shall now prove that
in a restricted sense this possibility may be real.

Let us consider the one-dimensional motion modeled by a particle of mass $m$
subject to a conservative force field plus the random effects of a vacuum
field. At the present stage of our reasoning what is matter is not the
physical origin of these random effects, but the way they change the particle
motion. The immediate impact falls over particle's energy whose fluctuations
are driven by this vacuum field. We describe this effect as
\[
H\left(  x\left(  t\right)  ,\dot{x}\left(  t\right)  \right)  =E_{0}%
+\mathcal{E}y\left(  t\right)
\]
where $\mathcal{E}$ is the noise source intensity and $y\left(  t\right)  $ is
the realization of a (dimensionless, zero mean) incomplete random variable
$Y$. Time discretization of this equation, with $\Delta t\equiv\varepsilon$,
leads to
\begin{equation}
\frac{m\left(  x_{n+1}-x_{n}\right)  ^{2}}{2\varepsilon^{2}}+V\left(
x_{n}\right)  -E_{0}=\mathcal{E}y_{n} \label{hamiltonian}%
\end{equation}
We are interested in a high noise intensity limit given by $\mathcal{E}%
\rightarrow\infty$, but subject to the condition that $\mathcal{E}\left\langle
y^{2}\right\rangle $ is finite. More specifically, we write $\mathcal{E}%
\equiv\alpha/\epsilon$ in such a way that $\left\langle H-E_{0}\right\rangle
=\mathcal{E}\left\langle y\right\rangle =0$ and
\begin{equation}
\sqrt{\left\langle \left(  H-E_{0}\right)  ^{2}\right\rangle }=\alpha
\frac{\sqrt{\left\langle y^{2}\right\rangle }}{\varepsilon}\equiv\frac{\alpha
}{\tau} \label{time-energy}%
\end{equation}
leading to uncertainty in the energy given by $\left(  \Delta E\right)
\tau\sim\alpha$. The number $\tau$ measures the characteristic time
fluctuations taking place during system evolution. Therefore, it must be
process-dependent. Its finiteness demands the limit of small fluctuations for
the variable $Y$. In order to use eqn$\left(  \ref{feymman1}\right)  $ for the
probability amplitude associated to this process we use normalized volume
integrals in this equation; thus, for fixed $\varepsilon$ , we do make the
substitution $dx\rightarrow\sqrt{\frac{m}{\varepsilon\alpha}}dx$ resulting in
the following expression for probability amplitudes
\begin{equation}
\psi\left(  x_{N}\right)  =\left(  \frac{m}{\varepsilon\alpha}\right)
^{N/2}\int\psi\left(  x_{0}\right)  \prod_{n=0}^{N-1}\sqrt{P\left(
\varepsilon\frac{\frac{m\left(  x_{n+1}-x_{n}\right)  ^{2}}{2\varepsilon^{2}%
}+V-E_{0}}{\alpha}\right)  }e^{i\varphi\left(  path\right)  }dx_{n}
\label{amplitude}%
\end{equation}
Considering the very small values of the random variable $Y$ we may limit
ourselves to first order terms in the expansion for the phase, so we have
$\varphi_{l}\left(  y_{0},....y_{N}\right)  =\varphi_{Nl}+\sum_{path}%
y+\mathcal{O}\left(  y^{2}\right)  $. We also scale energy reference level
choosing $E_{0}$ in such a way that $\sum\varphi_{Nl}=E_{0}/\mathcal{E}$, so
eqn$\left(  \ref{amplitude}\right)  $ leads to
\[
\psi\left(  x_{N}\right)  =\left(  \frac{m}{\varepsilon\alpha}\right)
^{N/2}\int\psi\left(  x_{0}\right)  \exp\left(  \frac{i\epsilon\sum_{path}%
H}{\alpha}+\mathcal{O}\left(  \epsilon^{2}\right)  \right)  \prod_{n=0}%
^{N-1}\sqrt{P\left(  \varepsilon\frac{\left(  H-E_{0}\right)  }{\alpha
}\right)  }dx_{n}%
\]
This linearization procedure hiddes the particular choice of the non-classical
therms present in the power series expansion for the phase. A more convenient
form for this equation comes out in the phase space representation, obtained
by the use of the following expression
\begin{equation}
\exp\left(  i\frac{m\left(  x_{n+1}-x_{n}\right)  ^{2}}{2\varepsilon\alpha
}\right)  =\sqrt{\frac{i\varepsilon}{2\pi m\alpha}}\int\exp\left(
\frac{i\varepsilon}{\alpha}\left(  \frac{p_{n}^{2}}{2m}-\frac{\left(
x_{n+1}-x_{n}\right)  }{\varepsilon}p_{n}\right)  \right)  dp_{n}
\label{Heisenb}%
\end{equation}
which after insertion on the equation for $\psi\left(  x_{N}\right)  $ and
discarding second order terms in $\epsilon$ results in the following naive
expression
\begin{equation}
\psi\left(  x_{N}\right)  =\left(  \frac{i}{2\pi}\right)  ^{N/2}\int
\psi\left(  x_{0}\right)  \exp\left(  \frac{-i}{\alpha}\sum_{path}%
\epsilon\mathcal{L}\right)  \prod_{n=0}^{N-1}\sqrt{P\left(  \varepsilon
\frac{\left(  H-E_{0}\right)  }{\alpha}\right)  }\frac{dx_{n}dp_{n}}{\alpha}
\label{FKac}%
\end{equation}
where $-\mathcal{L\equiv}\frac{p_{n}^{2}}{2m}+V\left(  x_{n}\right)
-\frac{\left(  x_{n+1}-x_{n}\right)  }{\varepsilon}p_{n}=H-p\dot{x}$ has to be
interpreted as the (''phase space'') particle's Lagrangean. Phase space
Lagrangean has a dual interpretation: the kinetic energy term is partially
determined by the velocity $\frac{\left(  x_{n+1}-x_{n}\right)  }{\varepsilon
}$ and partially by the independent momentum variable $p$. Application of
Euler-Lagrange equation to this Lagrangean selects the classical path and
gives directly the set of dynamical equations for the system
\begin{align}
p\left(  t\right)   &  =m\dot{x}\left(  t\right) \label{motion1}\\
\dot{p}\left(  t\right)   &  =-\frac{dV}{dx}\nonumber
\end{align}
In our stochastic model noise decouples momentum and velocity so all orbits in
phase space are now permitted. No deterministic relation between momentum and
velocity, like that shown in eqn$\left(  \ref{motion1}\right)  $, exists any
more although the velocity still is real and given by the time derivative of
the position. The same effect is verified when the motion is subject to a
Wiener noise and a Fokker-Planck equation, using the Ito approach, is
constructed. In this case it is possible to show that the obtained
Fokker-Planck equation admits a probability amplitude formalism which results
from a perturbation series on a variable conjugate to the momentum. The lowest
order terms are fully compatible with a quantum dynamics for the particle
\cite{fp}.

The above developments shows that space and momentum variables are related by
the unitary transformation of the eqn$\left(  \ref{Heisenb}\right)  $;
consequently they satisfy an equal time ''uncertainty principle'' since from
known properties of Fourier transform we must have
\[
\frac{\varepsilon}{\alpha}\left(  \Delta\left[  \frac{\left(  x_{n+1}%
-x_{n}\right)  }{\varepsilon}\right]  \Delta\left[  p_{n}\right]  \right)
=\frac{1}{\alpha}\left(  \Delta x_{n}\right)  \left(  \Delta p_{n}\right)
\geq2\pi
\]
valid for fixed $x_{n+1}$.

Probability amplitudes are then given by eqn$\left(  \ref{FKac}\right)  $
which is a phase-space Feynman-Kac formula in a most striking resemblance to
Quantum Mechanics. The difference is concentrated on the $\sqrt{P}$ term
associated to the still undefined bare probability $P$ and in the value of the
constant $\alpha$. In the limit of continuous time $\left(  \epsilon
\rightarrow0\right)  $ the condition of finite rms for the noise source, as
displayed in eqn$\left(  \ref{time-energy}\right)  $, demands that only
vanishingly small values of the random variable are relevant at infinitely
large noise intensity $\mathcal{E}$. This fact allows the substitution of
values for $P$ by its value at $y=0$ (equal to $P_{0}$), which is then
incorporated as a normalization constant in the probability amplitude. As a
result a finite probability distribution is obtained, since the lost
apodization induced by large fluctuations is compensated by this normalization
procedure. All these considerations complete a theory for amplitudes
calculated as
\begin{equation}
\psi\left(  x_{N}\right)  =\left(  \frac{iP_{0}}{2\pi}\right)  ^{N/2}\int
\psi\left(  x_{0}\right)  \exp\left(  \frac{-i\mathcal{A}_{path}}{\alpha
}\right)  \prod_{n=0}^{N-1}\frac{dx_{n}dp_{n}}{\alpha} \label{Feynm}%
\end{equation}
where $\mathcal{A}_{path}\mathcal{\equiv}\epsilon\sum_{path}\mathcal{L}$ is
the discrete time phase-space classical action and the value of $P_{0}$ being
incorporated in a proper Hilbert space normalization for $\psi\left(
x_{N}\right)  $. Notice that the bare probability structure is hidden in this
approximation which leads to a hidden process (or variable?) theory. An
important fact is that eqn$\left(  \ref{Feynm}\right)  $ is fully algorithmic
(e.g. using Monte Carlo) by the use of the prescriptions given here for path
space construction making trivial the mystery of how an equation like the
above one may naturally appear in theoretical models. The result we have
obtained is a Feynman-like path integral for a quantum particle; apart from a
normalization factor it seems that non-classical stochastic process, as
described here, is sufficiently rich to explain the probability amplitude
structure of one-dimensional Quantum Mechanics. The classical action being the
Onsager-Machlup functional \cite{onsager} for the underlying stochastic
process which acts not on probabilities but on amplitudes as required by the
non-classical nature of this random process. However the universal character
of the Planck's constant demands that the kind of noise we considered must
also be universal in order to set $\alpha=\hbar$ in the above reasoning and in
particular in the eqn$\left(  \ref{FKac}\right)  $, the generalized
Feynman-Kac formula. Saying differently, must exist a kind of noise field
capable to couple to any elementary particle wherever its physical character.
This is not a trivial task and go further in this direction now is premature
considering the primitive informations we have at hand besides the possibility
of incomplete nature for an existing fundamental stochastic vacuum. The
converse sense is also true. Feynman and Hibbs \cite{hibbs} showed that the
important paths for a quantum mechanical particle are those that are not
differentiable. In particular they have shown that $\left\langle \left(
x_{n+1}-x_{n}\right)  ^{2}/\epsilon^{2}\right\rangle \sim\epsilon^{-1}$ a
result compatible with fractal trajectories activated by a noise source. Thus
should exist, following the prescription given here, a non-classical random
variable associated to quantum processes, not available inside the limits of
the Quantum Theory because it should be the linearization of some hidden
incomplete stochastic process. What we have showed is how the hidden mechanism
(the HV mechanism) happens, masking this subjacent stochastic process and
leaving out only quantal effects, which shall depend on the linearized form of
the nonclassical terms through the $d$-terms set. Consequently it turns out
quite impossible distinguish Quantum Mechanics, at least for the simple case
treated here, from a incomplete random variable process.

As strange may appear the use of the phase space Lagrangean presents no
additional difficulties as well. In fact our model clearly defines the role of
each dynamical concept in quantum processes, specifying its character of
reality and their formal intrinsic relationship. We can advance further in
this reasoning if we look for a differential equation that solves the
amplitude problem. The first step is to find $\delta\psi=\psi\left(
x,t+\varepsilon\right)  -\psi\left(  x,t\right)  $ and owing to eqn$\left(
\ref{Feynm}\right)  $ this configures a moving boundary variational problem
since the end point of the action integral is changed. We have
\[
\delta\psi=\frac{-1}{i\alpha C}\int\psi\left(  x_{0}\right)  \left(
\delta\mathcal{A}\right)  \exp\left(  \frac{-i\mathcal{A}\left(  y,\dot
{y}\right)  }{\alpha}\right)  \mathcal{D}y\mathcal{D}p
\]
and to calculate $\delta\mathcal{A}$ the effect of changing the end point due
to particle's velocity must be considered. The result is \cite{elsgolc}
\[
\delta\mathcal{A=}\left(  \mathcal{L}-\dot{y}\left(  t\right)  \frac
{\partial\mathcal{L}}{\partial\dot{y}\left(  t\right)  }\right)  _{y\left(
t\right)  =x}dt+\frac{\partial\mathcal{L}}{\partial\dot{y}\left(  t\right)
}_{_{y\left(  t\right)  =x}}dx+\int^{t}\left(  \frac{\partial\mathcal{L}%
}{\partial y}-\frac{d}{dt^{%
%TCIMACRO{\UNICODE{0xb4}}%
%BeginExpansion
\acute{}%
%EndExpansion
}}\frac{\partial\mathcal{L}}{\partial\dot{y}}\right)  \delta y\left(  t^{%
%TCIMACRO{\UNICODE{0xb4}}%
%BeginExpansion
\acute{}%
%EndExpansion
}\right)  dt^{%
%TCIMACRO{\UNICODE{0xb4}}%
%BeginExpansion
\acute{}%
%EndExpansion
}
\]
where the second term in the coefficient of $dt$ corrects the time derivative
of the action due to particle's motion. From this equation we get the partial
derivatives of the amplitude:
\begin{align*}
\frac{\partial\psi}{\partial x}  &  =\frac{-1}{i\alpha C}\int\psi\left(
x_{0}\right)  p\exp\left(  \frac{-i\mathcal{A}}{\alpha}\right)  \mathcal{D}%
y\mathcal{D}p\\
\frac{\partial\psi}{\partial t}  &  =\frac{1}{i\alpha C}\int\psi\left(
x_{0}\right)  \left(  \frac{p^{2}}{2m}+V\left(  x\right)  \right)  \exp\left(
\frac{-i\mathcal{A}}{\alpha}\right)  \mathcal{D}y\mathcal{D}p
\end{align*}
Notice that information about particle's velocity is lost whilst the surviving
associated variable, the momentum, has no reality content; at most a
statistical interpretation of its value can be given. This view becomes more
evident if we note that
\[
\frac{\partial^{2}\psi}{\partial x^{2}}=\frac{-1}{\alpha^{2}C}\int\psi\left(
x_{0}\right)  p^{2}\exp\left(  \frac{-i\mathcal{A}}{\alpha}\right)
\mathcal{D}y\mathcal{D}p
\]
so a closed differential equation is obtained for the amplitude
\[
i\alpha\frac{\partial\psi}{\partial t}=-\frac{\alpha^{2}}{2m}\frac
{\partial^{2}\psi}{\partial x^{2}}+V\left(  x\right)  \psi
\]
defining its Hilbert space operator structure. Observe that the probability
amplitude depends only on spatial variables and time. No information
concerning the velocity is present although a resemblance to the (classical)
momentum through spatial derivatives are permitted with some care allowing the
usual association of the operator $-\frac{\alpha^{2}}{2m}\frac{\partial^{2}%
}{\partial x^{2}}+V\left(  x\right)  $ to the classical Hamiltonian. Our
formalism show how this exactly happen and to what extend this association is
valid. However remember that this equation is valid for a specific noise type
and its universal validity, which permits the substitution $\alpha=\hbar$,
demands a non-trivial vacuum physics.

The present theory has a quite general range and for the specific mechanical
model we are considering the formalism of Quantum Mechanics, if applicable, is
just the simplest description, that using the lowest order in the phase and
noise probability distribution expansion. It seems reasonable that the weakest
nonlinear contribution may generate corrections to eqn$\left(  \ref{FKac}%
\right)  $ which lies outside Quantum Theory itself while still maintaining
the nonclassical character of the formalism. Expressing differently we able to
predict hipper-quantum phenomena, those explicitly depending of the Q-rules
(in the present mechanical model means $d$-dependent terms) but not explained
by the use of the simple Feynman formula. This new class of quantum phenomena
would be described by considering the second order correction of eqn$\left(
\ref{phase}\right)  $ given by the set of $B$ coefficients. Thus, at least in
principle, we know how generalizations to Quantum Theory may appear and how to
test if Quantum Theory is fundamental. Nobody makes no doubt about the
capabilities of Quantum Mechanics in explain non-classical word. However once
it is considered a fundamental theory all possible predictions of nonclassical
phenomena must lie within the range of its formalism. The nonlinear correction
to the phase enable us to test wether or not in fact it is fundamental as well
as hopefully predict new phenomena never yet considered. We can say more. In
the continuum $\left(  \epsilon\rightarrow0\right)  $ only vanishingly small
values of the fluctuating variable $Y$ will survive and in this case the
linear approximation is rigorously true if the kernel matrix $B_{nl}^{ml}$ as
well as all higher order phase coefficients are topologically dense to survive
in the continuum. This means that Quantum Mechanics may be, in fact, a truly
hidden-variable theory of first type \cite{belinfante} and any tentative of
find or detect its stochastic nature be definitively unfruitful. In this case
an evidence of hipper-quantum phenomena means that space-time continuity is
broken at some scale thus generating information about the underground vacuum
physics necessarily hidden in usual quantum processes.

On the contrary, if the weakest nonlinear term survives in the continuum the
generalized Feynman-Kac formula, eqn$\left(  \ref{Feynm}\right)  $, changes
to
\begin{align*}
\psi\left(  x,t\right)   &  =\frac{1}{C}\int\psi\left(  x_{0}\right)
\exp\left(  \frac{-i\mathcal{A}_{path}}{\alpha}\right)  .\\
&  .\exp\left(  \frac{i}{\alpha^{2}}\int^{t}B\left(  u\left(  \tau\right)
,\dot{u}\left(  \tau\right)  ,u\left(  \varsigma\right)  ,\dot{u}\left(
\tau\right)  \right)  H\left(  u\left(  \tau\right)  ,\dot{u}\left(
\tau\right)  \right)  H\left(  u\left(  \varsigma\right)  ,\dot{u}\left(
\varsigma\right)  \right)  d\tau d\varsigma\right)  \mathcal{D}u\mathcal{D}p
\end{align*}
so in order to derive a differential equation for the wave function in a
similar way developed above, a generalization of the moving boundary
variational calculus must be done. Keeping terms linear in $B$ this may be
done yet with some involved calculations. The result presents a correction to
Schr\"{o}edinger equation (linear in $B$) where the important fact is that
vacuum terms are present. The HV mechanism is broken. This involved questions
transcends, by its nature, the limits of the discussions we intend to present
in this work so we deserve for the future additional tracks on this line.

Concluding, observe that the rules we introduced here for a incomplete random
variable may, at first glance, be so strange as the axioms of Quantum Theory
are. But once those rules are algoritmic, we are able to test them from a
heuristic point of view since they must belong to the vacuum phenomenology
which, at least in principle, is accessible to experiments. Breakdown of
causality and determinism, as introduced here, is not a big problem too
because no physical principle demands these reasonable assumptions be valid
outside our common sense perception. The main result of this work still is the
fact that a probability amplitude formalism is possible in describing the
class of stochastic process we introduced here. This is a general result which
may include physical processes as well. In this case a probability amplitude
description of the motion of a particle is obtained, with properties very
similar to those Quantum Theory makes use of. The price paid is the need of an
universal vacuum field in order to explain the generality of Quantum
Mechanics. At the present stage of our formulation we are not able to predict
all properties this vacuum field must possess. The apparent advantage over
existing hidden-variable theories is the algorithmic procedure and the
explicit demonstration of the HV mechanism. However it is important to cite
that we are not worried about justifications to Quantum Theory. If Nature
truly admits incomplete random variables in its basic realm Quantum Theory
shouldn't be considered as fundamental because in this case it cannot capture
the basic processes physical world possesses. Its success would be consequence
of fortuit epistemological and practical rules the HV mechanism enables. This
way we believe the present work opens tips to looking for new phenomenology in
this field. It should be indispensable, within the present context, a serious
investigation on vacuum properties, powered by an incomplete stochastic random
field, as a proper source for quantum behavior in Nature.


\bigskip

\begin{thebibliography}{99}
\bibitem{miloni}A. Einstein and L. Hopf, Ann. d. Phys. \textbf{33}, 1105
(1910), A. Einstein and O. Stern, Ann. d. Phys. \textbf{40}, 551 (1913), P. W.
Milonni, \textit{The Quantum Vacuum} (Academic Press, San Diego, 1994).

\bibitem{note}In fact a game demands more than one (competiting) player so
what we have defined is properly a lottery. A game concerns a multivariate
process. We do not treat this case in the present work because the main ideas
we intend show here are not affected by the number of players and can be
easily implemented for a true game. Nevertheless we still keep game's terminology.

\bibitem{kampen}N. G. van Kampen, \textit{Stochastic Processes in Physics and
Chemistry} (North-Holland, Amsterdan, 1981).

\bibitem{onsager}L. Onsager and S. Machlup, Phys. Rev. \textbf{91}, 1505 (1953)

\bibitem{graham}R. Graham, Z. Physik B \textbf{26}, 281 (1977)

\bibitem{colored}See for example Pesquera et all, Phys. Lett. \textbf{94A} 287
(1983), Wio et all, Phys. Rev. A \textbf{40}, 7312 (1989), Lehmann et all,
Phys. Rev E \textbf{62}, 6282 (2000)

\bibitem{gardner}C. W. Gardiner, \textit{Handbook of Stochastic Methods}
\ (Springer-Verlag, Berlin, 1990)

\bibitem{game}Drew Fudenberg and Jean Tirole, \textit{Game Theory} (Cambridge,
Mass., 1991)

\bibitem{fp}M. S. Torres Jr. and J. M. A. Figueiredo, quant-ph/0204123 (2002)

\bibitem{hibbs}R. P. Feynman and A. R. Hibbs, \textit{Quantum Mechanics and
Path Integrals} (MacGraw-Hill, 1965)

\bibitem{belinfante}F. J. Belinfante, \textit{A survey on Hidden-Variables
Theory}, (Pergamon Press, 1973)

\bibitem{elsgolc}L. E. Elsgolc, \textit{Calculus of Variations} (Pergamon
Press, Mass., 1962)
\end{thebibliography}
\end{document}